\documentclass[11 pt,oneside,onecolumn,a4paper]{article}
\usepackage{amsmath}
\usepackage{graphicx}
\DeclareGraphicsExtensions{.eps,.ps,.pdf}
\usepackage{bbm}
\usepackage{bm}
\usepackage{epsfig}
\usepackage[T1]{fontenc}
\usepackage{esint}

mm \leftmargin 5pt \rightmargin 5pt \hoffset= -15mm

\title{Damped oscillations of the energy of a bosonic bath due to spectral gaps and special initial correlations}
\author{Filippo Giraldi}

\date{\small{School of Chemistry and Physics, University of KwaZulu-Natal\\ 
and National Institute for Theoretical Physics (NITheP)\\
Westville Campus, Durban 4000, South Africa
\vspace{1em}\\Gruppo Nazionale per la
Fisica Matematica (GNFM-INdAM)\\
c/o Istituto Nazionale di Alta Matematica Francesco Severi\\
Citt\'a Universitaria, Piazza Aldo Moro 5, 00185 Roma, Italy}}

\begin{document}

\maketitle

 \begin{abstract}
The energy of the bosonic bath and the flow of quantum 
information are analyzed over short and long times in local dephasing channels for special correlated or factorized initial conditions, respectively, which involve thermal states. The continuous distribution of frequency modes of the bosonic bath exhibits a spectral gap over low frequencies. The bath energy shows oscillatory behaviors around the asymptotic value and information is alternatively lost and gained by the open system. Due to the low-frequency gap, the damped oscillations become regular over long times and the frequency of the oscillations coincides with the upper cut-off frequency of the spectral gap. Sequences of long-time intervals are obtained over which the bath energy increases (decreases), for the correlated initial conditions, and information is lost (gained) by the open system, for the factorized initial configurations, even at different temperatures. Such long-time correspondence between the variations of the bath energy and of the information is reversed if compared to the one obtained without the low-frequency gap. The correspondence fails if the spectral density is tailored according to power laws with odd natural powers near the upper cut-off frequency of the spectral gap.
\end{abstract}
\vspace{-0 mm}
 \maketitle


\section{Introduction}
\label{1}

In recent years, the vast research work on open quantum systems and quantum information processing, has encouraged to explore the relations between transfer of energy and flow of quantum information \cite{W,BP,MQTP}. The exchange of energy between open system and external environment exhibits various connections with non-Markovianity \cite{EnergyFlow1,EnergyFlow2,EnergyFlow3}. In a three-level system, the coupling with two non-Markovian bosonic baths at zero temperature generates an unidirectional flow of energy, over finite time intervals, from the non-Markovian to the more Markovian bath \cite{EnergyFlow3}. Non-Markovian evolution of open quantum systems can be interpreted via the trace distance measure as a flow of quantum information from the external environment back to the open system \cite{BnnMarkovPRL2009,nnMarkovNeg2LZPRA2011,FPRA2013}. Similarly to information backflow, the energy can flow from the environment back in the open system. Especially, energy backflow can be observed in the non-Markovian regime along with information backflow \cite{EnergyFlow1}. 

Controlling the flow of energy and of information in open quantum systems is a fundamental goal that can help in the realization of quantum information processing. Local dephasing channels constitute a referential model for the study of the flow of quantum information 
\cite{MPRAr2013,MPRA2014,QbtMPRA2014,MNJP2015,YPRB2003}. 
 In local dephasing channels the appearance of information backflow is determined by the spectral properties of the system \cite{MPRAr2013,MPRA2014}. The information flow can be controlled and reversed by acting on the interaction between open system and environment via dynamical decoupling pulses \cite{MNJP2015}. For ohmic-like environments the long-time flow of information exhibits a straightforward dependence on the low-frequency structure of the spectral density (SD) via the ohmicity parameter \cite{GPRA2017}. A simple and effective way of generating information backflow in local dephasing channels at zero temperature is the realization of a low-frequency gap in the continuous distribution of frequency modes \cite{GgapsXiv2017}. In fact, sequences of regular long-time intervals over which information backflow appears are generated by the low-frequency gap. In local dephasing channels the qubit experiences pure dephasing and no dissipation of energy. Consequently, no energy flows between the open system and the external environment. Still, the non-equilibrium energy of the bosonic bath varies and can be estimated via the SD for special correlated initial conditions \cite{M1,M2,M3}. In such cases the bath energy exhibits a straightforward connection with the information flow over long times \cite{GOSID2017}. In fact, in the super-ohmic regime and over long times, the spectral properties that induce increasing (decreasing) bath energy with the special correlated initial conditions, provide gain (loss) of information in the open system for factorized conditions, even for different values of the temperatures which are involved in the initial configurations \cite{GOSID2017}. As a continuation of the scenario described above, here, we intend to study the effect of a spectral gap on the time evolution of the bath energy and possible relations with the information which is gained or lost by the open quantum system.

The paper is organized as follows. Section \ref{2} is devoted to the description of the model and of the initial conditions. The non-equilibrium energy of the bosonic bath is introduced in Section \ref{3}. The SDs with the low-frequency gaps under study are defined in Section \ref{4}. The short- and long-time behaviors of the bath energy are analyzed in Section \ref{5}. The study of the flow of quantum information is performed in Section \ref{6}. Correspondences between variations of the bath energy and the information, which is gained or lost by the open system, are analyzed in Section \ref{7}. A summary of the results are provided, along with conclusions, in Section \ref{8}. Details of the calculations are given in the Appendix.

\section{Model and initial conditions}
\label{2}

A qubit (two-level system) is linearly coupled to a reservoir of field modes \cite{BP,W,MPRAr2013,MPRA2014,QbtMPRA2014}. The microscopic Hamiltonian $H$ of the whole system reads $H=H_S+H_{SE}+H_E$, where $H_S$ is the Hamiltonian of the qubit, $H_E$ the Hamiltonian of the bosonic environment, and $H_{SE}$ is the coupling term,
\begin{equation}
\hspace{-3em}H_S= \omega_0 \sigma_z, \hspace{1em} H_{SE}=\sum_k \sigma_z \left(g_k b_k+g^{\ast}_k b^{\dagger}_k\right), \hspace{1em} H_{E}= \sum_k \omega_k b^{\dagger}_k b_k.
 \label{H}
\end{equation}
The Planck and Boltzmann constants are equal to unity, $\hbar=k_B =1$, in the chosen system of units. The parameter $\omega_0$ refers to the transition frequency of the qubit, $\omega_k$ represents the frequency of the $k$th mode, while $b^{\dagger}_k$ and $b_k$ are the rising and lowering operator of the same mode, respectively. The coefficient $g_k$ represents the coupling strength between the qubit and the $k$th frequency mode. The index $k$ runs over the frequency modes. The operator $\sigma_z$ refers to the $z$-component of the Pauli spin operator \cite{W,BP}.

 The mixed state of the qubit at the time $t$,
is described by the reduced density matrix $\rho(t)$, which is obtained by tracing the density matrix of the whole system, at the time $t$, over the Hilbert space of the external environment \cite{BP}. If the state of the qubit is initially factorized from the thermal state of the bosonic bath the reduced density matrix evolves in the interaction picture according to the master equation \cite{RH1,RH2,RH3}
\begin{equation}
\dot{\rho}(t)=\gamma(t)\left(\sigma_z \rho(t) \sigma_z -\rho(t)\right).  \label{Eq1}
\end{equation}
The function $\gamma(t)$ is named dephasing rate and depends on the SD of the system and on the temperature of the initial thermal state of the bosonic bath. At zero temperature, $T=0$, the dephasing rate is referred here as $\gamma_0(t)$ and reads
\begin{eqnarray}
&&\gamma_0(t)=\int_0^{\infty} \frac{J\left(\omega\right)}{\omega}\,
\sin \left(\omega t \right) \, d\omega.  \label{gamma0} 
\end{eqnarray}
The function $J\left(\omega\right)$ is the SD of the system and depends on the coupling constants $g_k$ as follows,
\begin{equation}
J\left(\omega\right)=\sum_k \left|g_k\right|^2 \delta \left(\omega-\omega_k\right). \label{SD}
\end{equation}
At non-vanishing temperatures, $T>0$, the dephasing rate is represented here as $\gamma_T(t)$ and reads
\begin{equation}
\gamma_T(t)=\int_0^{\infty} \frac{J_T\left(\omega\right)}{\omega}\,\sin \left(\omega t\right) d\omega,  \label{gammaT}
\end{equation}
where the effective SD $J_T\left(\omega\right)$ is defined as
\begin{eqnarray}
&&\hspace{-0em}J_T\left(\omega\right)=J\left(\omega\right) \coth \frac{ \omega}{2 T}, \label{JT}
\end{eqnarray}
for every non-vanishing value of the temperature $T$.

The model under study provides a pure dephasing process of the qubit \cite{RH1,RH2,RH3}. The quantum coherence between the states $|0\rangle$ and $|1\rangle$ of the qubit is described by the off-diagonal element $\rho_{0,1}(t)$ of the density matrix that undergoes the following evolution \cite{RH1,RH2,RH3}, 
\begin{equation}
\rho_{0,1}(t)=\rho^{\ast}_{1,0}(t)=\rho_{0,1}(0)\, \exp \left\{-\Xi(t)\right\}.  \label{rho01t}
\end{equation}
The function $\Xi(t)$ represents the dephasing factor and depends on the temperature $T$ of the thermal bath and on the coupling between the system and the environment. At zero temperature, $T=0$, the dephasing factor is indicated here as $\Xi_0(t)$ and results in the following form,
\begin{eqnarray}
&&\Xi_0(t)=\int_0^{\infty} J\left(\omega\right)\, \frac{1- \cos\left(\omega t \right) }{\omega^2}\, d\omega.  \label{Xi0t}
\end{eqnarray}
If the external environment is initially in a thermal state, $T>0$, the dephasing factor is represented here as $\Xi_T(t)$ and reads
\begin{eqnarray}
\Xi_T(t)=\int_0^{\infty}\frac{J_T\left(\omega\right)}{\omega^2} \left(1-\cos\left(\omega t\right)\right). \label{XiTt}
\end{eqnarray}
Both for vanishing and non-vanishing temperature, the dephasing factor is related to the dephasing rate via the time derivative, $\gamma_0(t)=\dot{\Xi}_0(t)$ and $\gamma_T(t)=\dot{\Xi}_T(t)$.
According to Eq. (\ref{rho01t}), recoherence corresponds to negative values of the dephasing rate.

\subsection{Special correlated initial conditions} \label{21}

We focus on special correlated initial conditions which are obtained from the thermal equilibrium of the whole system at temperature $T$. Following Refs. \cite{M1,M2,M3}, the density matrix of the whole system is $\exp\left(-H/T\right)/Z_0^{\prime}$, where $Z_0^{\prime}$ is the normalization constant, $Z_0^{\prime}=\operatorname{Tr} \left\{\exp \left(-H/T\right)\right\}$. The symbol $\operatorname{Tr}$ denotes 
the trace operation over the Hilbert space of the whole system. A selective measurement \cite{measure1,measure2} induces the qubit in a pure state $|\phi_0\rangle$ and the whole system in the state $P_0 \exp\left(-H/T\right) P_0/Z_0$. The selective measurement is described by the projector operator $P_0$, given by $P_0=\left|\phi_0\rangle\langle\phi_0\right|$, while $Z_0$ is the normalization constant, $Z_0=\operatorname{Tr} \left\{P_0 \exp\left(-H/T\right) P_0\right\}$. 
In this way, the initial condition of the whole system is 
\begin{eqnarray}
\rho(0)=\left|\phi_0\rangle\langle\phi_0\right|\otimes \rho_E(0).
\label{rho0corr}
\end{eqnarray}
The mixed state $\rho_E(0)$ represents the initial state of the bosonic bath and reads \cite{M1,M2,M3}
\begin{eqnarray}
\rho_E(0)=\frac{\langle \phi_0\left|\exp\left(- H/T\right) \right|\phi_0\rangle}{\operatorname{Tr}_{E}\langle \phi_0\left|\exp\left(- H/T\right) \right|\phi_0\rangle}. \label{rhoE0}
\end{eqnarray}
The symbol $\operatorname{Tr}_{E}$ refers to the trace operation over the Hilbert space of the external environment. The mixed state $\rho_E(0)$ of the bosonic bath depends on the state $|\phi_0\rangle$ of the qubit and on the interaction Hamiltonian $H_{SE}$. Consequently, in such initial configuration the qubit and thermal bath are correlated. Refer to 
\cite{M1,M2,M3,measure1,measure2,measure3,measure4,measure5,measure6} for details.

\section{The energy of the bosonic bath}\label{3}

The non-equilibrium energy of the bosonic bath is evaluated as the sum over the bosonic modes of the expectation values of the number of excitations \cite{M1,M2,M3},
\begin{equation}
\epsilon_E(t)=\sum_k \omega_k n_k(t), \label{EnDef}
\end{equation}
where $n_k(t)=\operatorname{Tr}\left\{\rho(0) b^{\dagger}_k(t)b_k(t)\right\}$. The 
operators $b^{\dagger}_k(t)$ and $b_k(t)$ represent the rising and lowering operators of the $k$th frequency mode in the Heisemberg picture, respectively, at time $t$. The index $k$ runs over the bosonic modes. The whole system is prepared in the special correlated initial condition $\rho(0)$, given by $\rho(0)=\left|\phi_0\rangle\langle\phi_0\right|\otimes \rho_E(0)$. where the initial state of the environment, $\rho_E(0)$, is given by Eq. (\ref{rhoE0}). Following Ref. \cite{M2}, if the whole system is initially prepared in the mentioned initial condition the non-equilibrium energy of the bosonic bath is
\begin{eqnarray}
\epsilon_E(t)=\epsilon_E(0)+d_0 \left(\eta_{1}-\Pi(t)\right). \label{Et}
\end{eqnarray}
For a discrete distribution of frequency modes the initial bath energy is given by the expression \cite{M2} 
\begin{eqnarray}
\epsilon_E(0)=\sum_k \frac{\omega_k}{\exp\left( \omega_k/T\right)-1}+ \eta_{1}, \label{E0}
\end{eqnarray}
while for a continuous distribution of frequency modes the initial bath energy reads
\begin{eqnarray}
\epsilon_E(0)=\int_0^{\infty} \frac{\omega\, r\left(\omega\right)}{\exp\left( \omega/T\right)-1}\, d \omega+ \eta_{1}. \label{E0c}
\end{eqnarray}
 The parameter $\eta_{1}$ is defined in terms of the SD both for a discrete and a continuous distribution of frequency modes as
\begin{eqnarray}
\eta_{1}=\int_0^{\infty}\frac{J\left(\omega\right)}{\omega} \,d\omega, \label{eta1}
\end{eqnarray}
while the function $r\left(\omega\right)$ denotes the density of the bosonic modes \cite{BP,W} at frequency $\omega$.
The parameter $d_0$ is given by the form \cite{M2}
$$\hspace{-1em}d_0=2 \left(1+ \langle\ \phi_0\left|\sigma_3\right|\phi_0 \rangle
\frac{\sinh\left( \omega_0/T\right)
-\langle\ \phi_0\left|\sigma_3\right|\phi_0 \rangle\cosh\left( \omega_0/T\right)}
{\cosh\left( \omega_0/T\right)-\langle\ \phi_0\left|\sigma_3\right|\phi_0 \rangle\sinh\left( \omega_0/T\right)}\right).
$$
The parameter $d_0$ vanishes for $\left|\langle\ \phi_0\left|\sigma_3\right|\phi_0 \rangle\right|=1$, and is positive for\\ $\left|\langle\ \phi_0\left|\sigma_3\right|\phi_0 \rangle\right|<1$. Consequently, the energy of the bosonic bath is 
constant, $\epsilon_E(t)=\epsilon_E(0)$, for $\left|\langle\ \phi_0\left|\sigma_3\right|\phi_0 \rangle\right|=1$, while it is time-dependent for \\$\left|\langle\ \phi_0\left|\sigma_3\right|\phi_0 \rangle\right|<1$. The function $\Pi(t) $ reads \cite{M2}
\begin{eqnarray}
&&\hspace{-3em}\Pi(t)=\int_0^{\infty}\frac{J\left(\omega\right)}{\omega}\, \cos\left(\omega t\right)\, d \omega,
\label{Pit}
\end{eqnarray}
and drives the evolution of the bath energy via Eq. (\ref{Et}). This function can be studied in terms of the SD of the system by following the analysis of the bath correlation function which is performed in Ref. \cite{GXiv2016}. The SDs under study are described below for the sake of clarity.

\section{Spectral gaps and spectral densities}\label{4}

Spectral gaps in the distribution of frequency modes can be created via periodic dielectric structures which characterize photonic band gap materials \cite{PBGLNNBRPP2000,KJMO1994,PCJbook,PBGYJP1993,PBGJWPRB1991,PBG1}. Let $\omega_g$ be the upper cut-off frequency of the spectral gap, the density of frequency modes $\rho\left(\omega\right)$ diverges for $\omega\to\omega_g^{+}$ approximately proportional to the form $
\left(\omega-\omega_g\right)^{\left(1-d_0\right)/2}\Theta\left(\omega-\omega_g\right)$, where $\Theta(\omega)$ is the Heaviside step function and $d_0$ is the dimension of the surface of the Brillouin zone that is spanned by the band edge modes with vanishing group velocity. The dimension $d_0$ is larger than unity in photonic band gap materals. In the band-gap edge, $\omega \to \omega_g^+$, the SD follows the inverse-power-law diverge of the density of modes, if the coupling function varies slowly at the edge of the gap. Refer to \cite{PBGLNNBRPP2000,KJMO1994,PCJbook,PBGYJP1993,PBGJWPRB1991,PBG1} for details.

Due to the spectral gap, the SDs under study vanish below the edge of the low-frequency gap, $J\left(\omega\right)=0$ for $0\leq \omega<\omega_g$. The adoption of a continuous distribution of frequency modes requires the constraint \cite{SDPlenio1,ReedSimonBook}
\begin{eqnarray}
\int_{\omega_g}^{\omega_M} \frac{J\left(\omega\right)}{\omega}\, d\omega<\infty. \label{ConstrSD0}
\end{eqnarray}
The maximum mode frequency $\omega_M$ can be either finite or infinite. For $\omega\geq \omega_g$ the SDs are characterized by the dimensionless auxiliary function $\Omega\left(\nu\right)$, which is defined by the scaling property $J \left( \omega_g+ \omega_s\nu \right)=\omega_s\Omega\left(\nu \right)$ for every $\nu\geq 0$, where $\omega_s$ is a typical scale frequency of the system. The two general classes of SDs under study have been introduced in Ref. \cite{GgapsXiv2017} and are tailored near the cut-off frequency $\omega_g$ according to power laws. The integrable inverse power laws refer to the divergences of photonic band gap materials. Powers of logarithmic forms are included in the power series expansions as possible perturbations of the power-law profiles. Such logarithmic powers are natural valued for the first class of SDs, and are arbitrarily real for the second class. See Ref. \cite{GgapsXiv2017} for details. For the sake of clarity, the definitions of the two classes of SDs are reported below.

\subsection{First class of spectral densities}\label{41}

The auxiliary functions $\Omega\left(\nu\right)$ which define the first class of SDs are continuous for every $\nu> 0$ and behave for $\nu\to 0^+$ as \cite{BleisteinBook}
\begin{eqnarray}
&&\hspace{-0em}\Omega\left(\nu\right)\sim 
\sum_{j=0}^{\infty}
\sum_{k=0}^{n_j}c_{j,k} \nu^{\alpha_j}\left(- \ln \nu\right)^k,  \label{o0log} 
\end{eqnarray}
where $\alpha_0>-1$, $\infty> n_j\geq 0$, $\alpha_{j+1}>\alpha_j$ for every $j\geq 0$, and $\alpha_j\uparrow +\infty$ as $j\to +\infty$. The coefficients $c_{j,k}$ must be chosen in such a way that the auxiliary functions $\Omega\left(\nu\right)$ are 
non-negative. Consequently the coefficient $c_{0,n_0}>0$ must be positive, $c_{0,n_0}>0$. The low-frequency behavior (\ref{o0log})
provides dominant power laws for $n_0=0$, and the integrable inverse-power-law divergences, appearing for $0 >\alpha_0>-1$, are integrable. In case $\alpha_0=0$, the logarithmic power $n_0$ must vanish due to the constraint of summability. For $\alpha_0>0$, the logarithmic singularity in $\nu=0$ can be removed by setting $\Omega(0)=0$. If support of the SDs is infinite, $\omega_{\textrm{max}}=\infty$, 
the integrability of the SD and the constraint (\ref{ConstrSD0}) are guaranteed by requiring $\Omega\left(\nu\right)= \mathcal{O}\left(\nu^{-1-\chi_0}\right)$. Sufficiently fast decays are required as $\left|\operatorname{Im} \,s \right|\to +\infty$ for the Mellin transform, and the analytic continuation \cite{BleisteinBook,Wong-BOOK1989}, of auxiliary function $\Omega\left(\nu\right)$ and of the functions $\Lambda_0\left(\nu\right)$ and $\Lambda_T\left(\nu\right)$, which are defined in the Appendix. Refer to the Appendix and to \cite{GXiv2016} for details.

\subsection{Second class of spectral densities}\label{42}

The auxiliary functions which define the second class of SDs behave for $\nu\to 0^+$ as
\begin{eqnarray}
&&\hspace{-0em}\Omega\left(\nu\right)\sim \sum_{j=0}^{\infty}w_j
\, \nu^{\alpha_j} \left(-\ln \nu\right)^{\beta_j}. \label{OmegaLog0}
\end{eqnarray}
The powers $\alpha_j$ fulfill the constraints that are reported in the previous Subsection, while the logarithmic powers $\beta_j$ are arbitrarily real, either positive or negative, or vanishing. Again, if the logarithmic power $\beta_0$ vanishes, the asymptotic expansion (\ref{OmegaLog0}) provides dominant
 power laws and integrable inverse-power-law divergences for $0 >\alpha_0>-1$. In case $\alpha_0=0$, the logarithmic power $\beta_0$ must vanish due to the constraint of summability. The constraint (\ref{ConstrSD0}) is fulfilled. Since the auxiliary functions $\Omega\left(\nu\right)$ are non-negative, the coefficients $w_j$ must be chosen accordingly, and the constraint $w_0>0$ is required. Again, for $\alpha_0>0$ the logarithmic singularity in $\nu=0$ is removable by defining $\Omega(0)=0$. Arbitrarily small, positive (negative) values of the first logarithmic power $\beta_0$ result in arbitrarily small increases (decreases) in the power-law profiles which descibe the SD near the edge of the spectral gap. The functions $\Omega\left(\nu\right)$ are rrequired to be continuous and differentiable over the support and summable. 

 Let $\bar{n}$ be the least natural number such that $\bar{n}\geq\alpha_{\bar{k}}$, where $\alpha_{\bar{k}}$ is the least of the powers $\alpha_k$ which are larger than or equal to unity. The function $\Omega^{\left(\bar{n}\right)}\left(\nu\right)$ is defined as the $\bar{n}$th derivative of the auxiliary function and must be continuous on the interval $\left(0,\infty\right)$. The integral $\int_0^{\infty}\Omega\left(\nu\right)\exp\left(-\imath \zeta \nu\right) d \nu$ must converge uniformly for all sufficiently large values of the variable $\zeta$ and the integral $\int \Omega^{\left(\bar{n} \right)}\left(\nu\right)\exp\left(-\imath \zeta \nu\right) d \nu$ has to converge at $\nu=+\infty$ uniformly for all sufficiently large values of the variable $\zeta$. The auxiliary functions must be differentiable $k$ times and the corresponding derivatives are required to fulfill as $\nu\to 0^+$ the asymptotic expansion
$$\Omega^{(k)}\left(\nu\right)\sim \sum_{j=0}^{\infty}w_j
\, \frac{d^k}{d\nu^k}\left(\nu^{\alpha_j} \left(-\ln \nu\right)^{\beta_j}\right),$$
for every $k=0,1, \ldots,\bar{n} $, where $\bar{n}$ is the non-vanishing natural number defined above. Furthermore, for every $k=0, \ldots,\bar{n}-1$, the function $\Omega^{(k)}\left(\nu\right)$ must vanish in the limit $\nu\to +\infty$. The above constraints are based on the asymptotic analysis performed in Ref. \cite{WangLinJMAA1978}.

The comparison between the first and second class of SDs under study, suggests that the logarithmic powers appearing in the definition of the first class of SDs, are natural valued. Instead, the logarithmic powers adopted in the definition of the second class of SDs, are real valued. Such arbitrariness requires more constraints but allows the perturbation of the power-law profiles which describe the SDs near the upper cut-off frequency $\omega_g$ of the spectral gap. See Refs. \cite{GgapsXiv2017,GXiv2016,GPRA2017,GOSID2017} for details. In both the classes under study, the SDs are non-negative, bounded and summable, due to physical grounds and, except for these constraints, are arbitrarily tailored over high frequencies.

\section{Damped oscillations of the bath energy}\label{5}

If the distribution of frequency modes of the bosonic bath is continuous and exhibits a spectral gap over low frequencies with upper cut-off frequency $\omega_g$, the energy of the bosonic bath shows oscillations over the time scale $1/\omega_g$. The oscillations become damped over long times, $t \gg 1/\omega_s$, and the energy tends to the asymptotic value. The features of the oscillations and of the damping are determined by the structure of the SD near the cut-off frequency $\omega_g$. These behaviors are descried analytically below.

Due to the presence of the spectral gap, the bath energy is obtained from Eqs. (\ref{Et}), (\ref{E0c})-(\ref{Pit}) by introducing $\omega_g$ as the lower extremum of integration. In this way, the bath energy results in the following form,
\begin{eqnarray}
\epsilon_E(t)=\epsilon_E\left(\infty\right)-d_0\Pi(t), \label{Etosc}
\end{eqnarray}
where $\epsilon_E\left(\infty\right)$ is the asymptotic value,
\begin{eqnarray}
\epsilon_E\left(\infty\right)=\epsilon_E(0)+d_0 \eta_{1}. \label{Einfty}
\end{eqnarray}
The damped oscillations are induced by the spectral gap via the function $\Pi(t)$, given by
 \begin{eqnarray}
\Pi(t)=\varphi_c(t) \cos \left(\omega_g t\right)-\varphi_s(t) \sin \left(\omega_g t\right). \label{Pitgapcs}
\end{eqnarray}
The functions $\varphi_c(t)$ and $\varphi_s(t)$ are defined 
in terms of the SD as below \cite{GgapsXiv2017},
\begin{eqnarray}
&&\hspace{-1em}\varphi_c(t) =\int_0^{\infty}\frac{J\left(\omega_g+ \omega^{\prime}\right)}{\omega_g+ \omega^{\prime}}\, \cos\left(\omega^{\prime} t\right) d \omega^{\prime}, \label{phc} \\
&&\hspace{-1em}\varphi_s(t) =\int_0^{\infty}\frac{J\left(\omega_g+ \omega^{\prime}\right)}{\omega_g+ \omega^{\prime}}\, \sin\left(\omega^{\prime} t\right) d \omega^{\prime}. \label{phs}
\end{eqnarray}
The expression (\ref{Pitgapcs}) suggests that the bath energy evolves over two time scales, $1/\omega_s$ and $1/\omega_g$. Both the time scales are determined by the SD which depends on the structure of the reservoir of field modes and on the coupling strength. If the condition $\varphi^2_c(t)+\varphi^2_s(t)>0$ holds for every $t \geq 0$, the function $\Pi(t)$, given by Eq. (\ref{Pitgapcs}), is equivalent to the form
\begin{eqnarray}
\Pi(t)=R_1(t) \cos \left(\omega_g t+\phi(t)\right).
 \label{Pitgap}
\end{eqnarray} The time-dependent angle $\phi(t)$ vanishes for $t=0$, i.e., $\phi(0)=0$, and is defined for $t>0$ via the functions $\varphi_c(t)$ and $\varphi_s(t)$ as below \cite{GgapsXiv2017},
\begin{eqnarray}
&&\hspace{-1em}\phi(t)=0,\hspace{1em}\text{if}\hspace{1em} \varphi_s(t)=0 \hspace{1em}\text{and}\hspace{1em}\varphi_c(t)>0,\label{phi0} \\
&&\hspace{-1em}\phi(t)=\pi,\hspace{1em}\text{if}\hspace{1em} \varphi_s(t)=0 \hspace{1em}\text{and}\hspace{1em}\varphi_c(t)<0,\label{phi1} \\
&&\hspace{-1em}\phi(t)=\operatorname{arccot} \frac{\varphi_c(t)}{\varphi_s(t)},\hspace{1em}\text{if}\hspace{1em} \varphi_s(t)>0,\label{phi2} \\
&&\hspace{-1em}\phi(t) =\pi +\operatorname{arccot} \frac{\varphi_c(t)}{\varphi_s(t)},\hspace{1em}\text{if}\hspace{1em} \varphi_s(t)<0. \label{phi3}
\end{eqnarray}
In general, the angle $\phi(t)$ is not a continuous function of time. On the contrary, the function $\phi(t)$ becomes continuous for every $t\geq 0$ and differentiable for every $t>0$ and bounded as $t\to \infty$ if the sine transform $\varphi_s(t)$ is positive (negative), $\varphi_s(t)>_{\left(<\right)}0$ for every $t > 0$, and if the cosine transform $\varphi_c(t)$ and the sine transform $\varphi_s(t)$ are continuous for every $t \geq 0$, and 
differentiable for every $t>0$. For example, the constraint \cite{GgapsXiv2017}
\begin{eqnarray}
J^{\prime}\left(\omega\right)<\frac{J\left(\omega\right)}{\omega}, \label{conSDphisp}
\end{eqnarray}
holding for every $\omega>\omega_g$, guarantees that the sine transform $\varphi_s(t)$ is positive, $\varphi_s(t)>0$ for every $t > 0$, and the angle $\phi(t)$, given by Eq. (\ref{phi2}), belongs to the interval $\left.\right]0,\pi\left[\right.$. 
The function $R_1(t)$ represents the damped time-dependent amplitude of the irregular oscillations, $R_1(t)=\sqrt{\varphi^2_c(t)+\varphi^2_s(t)}$, and vanishes over long times \cite{GgapsXiv2017}.

\subsection{Short- and long-time behavior of the bath energy}\label{}

The short-time behavior of the bath energy depends on integral and high frequency properties of the SD. Let the SD decays sufficiently fast over high frequencies: $\chi_0>1$ for the first class, or $\chi_0>3$ for the second class. Under such conditions, the bath energy increases quadratically in time for $t \ll \min \left\{1/\omega_g,1/\omega_s\right\}$,
\begin{eqnarray}
\epsilon_E(t)\sim\epsilon_E(0)+l_E t^2. \label{Etshort}
\end{eqnarray}
The parameter $l_E$ is defined in the Appendix in terms of the frequency $\omega_g$ and integral properties of the SD.

Over long times the damped oscillations of the bath energy are described by Eqs. (\ref{Etosc}) and (\ref{Pitgap}) and by the long-time decays of the functions $\varphi_c(t)$ and $\varphi_s(t)$. For both the classes of SDs under study, such long-time decays are given by logarithmic and power laws and are monotonic, either strictly increasing or decreasing, and non-vanishing. Refer to \cite{GgapsXiv2017} for details. A further description of the long-time oscillations is obtained in terms of the limit of the angle $\phi(t)$ for $t\to+\infty$. Since the sign of the sine transform $\varphi_s(t)$ is constant over long times, $t \gg 1/\omega_s$, the function $\phi(t)$ is continuous and the limit of the angle $\phi(t)$ for $t\to+\infty$ exists finite. Such limit has been studied in Ref. \cite{GgapsXiv2017} and depends on the structure of the SD near the cut-off frequency $\omega_g$ as below. Consider the two classes of SDs under study which are defined in Sect. \ref{4}. For the allowed negative values of the power $\alpha_0$, the limit $\phi\left(\infty\right)$ is given by the form
\begin{eqnarray}
&&\hspace{-2em}\phi\left(\infty\right)=
\frac{\pi}{2} \left(1+\alpha_0\right), \hspace{1em}\textrm{if} \hspace{1em} 
0>\alpha_0>-1. \label{phiinf0} 
\end{eqnarray}
 If the power $\alpha_0$ is positive and is not a natural number
 we find
\begin{eqnarray}
&&\hspace{-2em}\phi\left(\infty\right)=\pi\left(\frac{1+\alpha_0}{2}
-\lfloor\frac{1+\alpha_0}{2} \rfloor+\Theta\left(-\cos
\left(\frac{\pi}{2} \alpha_0\right)\right)\right),  \label{phiinf1} \\&&\hspace{-2em} 
\textrm{if} \hspace{1em}\alpha_0>0, \hspace{1em}\textrm{and} \hspace{1em}\alpha_0 \neq \lfloor \alpha_0\rfloor. \nonumber
\end{eqnarray}
If the power $\alpha_0$ takes even natural values the limit $\phi\left(\infty\right)$ is
\begin{eqnarray}
&&\hspace{-2em}\phi\left(\infty\right)=\frac{\pi}{2} \left(2-(-1)^{m}\right),
\hspace{1em}\textrm{if} \hspace{1em}\alpha_0=2 m,\hspace{1em}\textrm{and} \hspace{1em}m=0,1,2,\ldots. \label{phiinf2}
\end{eqnarray}
If the power $\alpha_0$ takes odd natural values and the logarithmic power $n_0$, for the first class of SDs, or $\beta_0$, for the second class, does not vanish, the limit $\phi\left(\infty\right)$ is given by
\begin{eqnarray}
&&\hspace{-2em}\phi\left(\infty\right)=\frac{\pi}{2} \left(3-(-1)^{m}\right),
\hspace{1em}\textrm{if} \hspace{1em}\alpha_0=1+2 m,\hspace{1em} n_0>0,\label{phiinf3}\\
&&\hspace{-2em}\beta_0\neq 0,\hspace{1em}\textrm{and} \hspace{1em}m=0,1,2,\ldots. \nonumber
\end{eqnarray}
If the power $\alpha_0$ takes odd natural values, $\alpha_0=1+2 m$ with $m=0,1,2,\ldots$, and the logarithmic power $n_0$, for the first class of SDs, or $\beta_0$, for the second class, vanishes, the limit $\phi\left(\infty\right)$ is determined by the power $\alpha_{k_1}^{\prime}$ appearing in the series (\ref{Ls01}) of the Appendix. The natural number $k_1$ the least non-vanishing index such that the power $\alpha^{\prime}_{k_1}$ is not an odd natural number or $\alpha^{\prime}_{k_1}=1+ 2m_1$, where $m_1$ is a natural number, and $n^{\prime}_{k_1}$, for the first class of SDs, or $\beta^{\prime}_{k_1}$, for the second class, is a non-vanishing natural number. If the power $\alpha_{k_1}^{\prime}$ is not an odd natural number we find
\begin{eqnarray}
&&\hspace{-2em}\phi\left(\infty\right)=
\frac{\pi}{2} \left(2+ \left((-1)^m-1\right)\operatorname{sign}\left(\cos\left(\frac{\pi}{2}\alpha^{\prime}_{k_1}\right)\right)\right),
\label{phiinf4}\\
&&\hspace{-2em}\textrm{if} \hspace{1em}
\alpha_0=1+2 m, \hspace{1em}\textrm{and} \hspace{1em} \,n_0=\beta_0= 0, \nonumber
\end{eqnarray}
while, if $\alpha_{k_1}^{\prime}=1+2 m_1$, the limit is 
\begin{eqnarray}
&&\hspace{-2em}\phi\left(\infty\right)=\frac{\pi}{2} \left(2
+(-1)^{m_1}\left((-1)^{m}-1\right)\right),\label{phiinf5}
\\&&\hspace{-2em}\textrm{if} \hspace{1em}\alpha_0=1+2 m,
\hspace{1em}\textrm{and} \hspace{1em}n_0=\beta_0= 0. \nonumber
\end{eqnarray}

 Since the limit $\phi\left(\infty\right)$ exists finite, the quantity $\left|\phi(t)-\phi\left(\infty\right)\right|$ vanishes for $t\gg 1/\omega_s$. We focus on long times $t$ which fulfill the constraint 
\begin{eqnarray}
\left|\phi(t)-\phi\left(\infty\right)\right|<\varepsilon_0, \label{Cepsilon0}
\end{eqnarray}
where $\varepsilon_0>0$. Over such long times we find 
\begin{eqnarray}
&&\epsilon_E\left(\infty\right)-\epsilon_E(t)<d_0 R_1(t)\left(\cos\left(\omega_g t+\phi\left(\infty\right)\right)+\varepsilon_0\right),\label{EtAsympt1}\\&&
\epsilon_E\left(\infty\right)-\epsilon_E(t)>d_0 R_1(t)\left(\cos\left(\omega_g t+\phi\left(\infty\right)\right)-\varepsilon_0\right).\label{EtAsympt2}
\end{eqnarray}
The above expressions show that the bath energy oscillates around the asymptotic values $\epsilon_E\left(\infty\right)$. The oscillations become arbitrarily close to the regular form $\cos\left(\omega_g t+\phi\left(\infty\right)\right)$ and the amplitude $d_0 R_1(t)$ tends to vanish over long times, as the positive parameter $\varepsilon_0$ becomes arbitrarily small. 

\subsection{Variations of the bath energy}\label{}

The variations of the bath energy are described by the sign of the time derivative $\dot{\epsilon}_E(t)$, which is given by the expression
\begin{eqnarray}
\dot{\epsilon}_E(t)=d_0 \int_{\omega_g}^{\infty} J\left(\omega\right) \sin\left(\omega t\right) d \omega. \label{dotE}
\end{eqnarray}
Again, the oscillations can be studied via the form
 \begin{eqnarray}
\dot{\epsilon}_E(t)=d_0\left(\theta_c(t) \sin \left(\omega_g t\right)+\theta_s(t) \cos \left(\omega_g t\right)\right). \label{dotEgap}
\end{eqnarray}
The functions $\theta_c(t)$ and $\theta_s(t)$ are defined 
in terms of the SD as 
\begin{eqnarray}
&&\hspace{-1em}\theta_c(t) =\int_0^{\infty}J\left(\omega_g+ \omega^{\prime}\right)\, \cos\left(\omega^{\prime} t\right) d \omega^{\prime}, \label{thetac} \\
&&\hspace{-1em}\theta_s(t) =\int_0^{\infty}J\left(\omega_g+ \omega^{\prime}\right)\, \sin\left(\omega^{\prime} t\right) d \omega^{\prime}. \label{thetas}
\end{eqnarray}
If the constraint $\theta^2_c(t)+\theta^2_s(t)>0$ holds for every $t \geq 0$, the time derivative of the bath energy is equivalent to the form
\begin{eqnarray}
\dot{\epsilon}_E(t)=d_0 R_2(t) \sin \left(\omega_g t+\psi(t)\right), \label{dotEgapS}
\end{eqnarray}
where $R_2(t)=\sqrt{\theta^2_c(t)+\theta^2_s(t)}$. Similarly to the definition of the angle $\phi(t)$, the time-dependent angle $\psi(t)$ vanishes for $t=0$, i.e., $\psi(0)=0$, and is defined for $t>0$ via the functions $\theta_c(t)$ and $\theta_s(t)$ as below,
\begin{eqnarray}
&&\hspace{-1em}\psi(t)=0,\hspace{1em}\text{if}\hspace{1em} \theta_s(t)=0 \hspace{1em}\text{and}\hspace{1em}\theta_c(t)>0,\label{psi0} \\
&&\hspace{-1em}\psi(t)=\pi,\hspace{1em}\text{if}\hspace{1em} \theta_s(t)=0 \hspace{1em}\text{and}\hspace{1em}\theta_c(t)<0,\label{psi1} \\
&&\hspace{-1em}\psi(t)=\operatorname{arccot} \frac{\theta_c(t)}{\theta_s(t)},\hspace{1em}\text{if}\hspace{1em} \theta_s(t)>0,\label{psi2} \\
&&\hspace{-1em}\psi(t) =\pi +\operatorname{arccot} \frac{\theta_c(t)}{\theta_s(t)},\hspace{1em}\text{if}\hspace{1em} \theta_s(t)<0. \label{psi3}
\end{eqnarray}
Again, in general, the angle $\psi(t)$ is not a continuous function of time. Instead, the function $\psi(t)$ is continuous for every $t\geq 0$, bounded for $t\to \infty$ and differentiable for every $t>0$ if the sine transform $\theta_s(t)$ is positive (negative), $\theta_s(t)>_{(<)}0$ for every $t > 0$, and if the cosine transform $\theta_c(t)$ and the sine transforms $\theta_s(t)$ are continuous for every $t \geq 0$, and differentiable for every $t>0$. For example, if the SD is decreasing and differentiable, $J^{\prime}\left(\omega \right)< 0$ for every $\omega>\omega_g$, the sine transform $\theta_s(t)$ is positive, $\theta_s(t)>0$ for every $t > 0$, and the angle $\psi(t)$, given by Eq. (\ref{psi2}), belongs to the interval $\left.\right]0,\pi\left[\right.$.

In general, the increasing or decreasing regime of the bath energy are not regular. Still, temporal bounds on such regimes can be determined if the angle $\psi(t)$ is a continuous  function of time and bounded for $t \to \infty$. Consider continuous and differentiable functions $\theta_c(t)$ and $\theta_s(t)$, and $\theta_s(t)>0$ for every $t > 0$. In this way, the function $\psi(t)$ is continuous and differentiable and belongs to the interval $\left.\right] 0,\pi\left[\right.$. According to Eq. (\ref{dotEgapS}), the first interval over which the bath energy increases, occurs for $t>0$ and has already ended at the instant $\pi/\omega_g$, and the energy decreases. In general, for every $n=2,3,\ldots$, the $n$th regime of increasing energy has already started at the instant $2(n-1)\pi/\omega_g$, and the energy increases, and has already ended at the instant $(2n-1)\pi/\omega_g$, and the energy decreases.

\subsection{Short- and long-time behaviors}\label{}

According to Eq. (\ref{Etshort}), the bath energy increases quadratically over short times, $t \ll \min \left\{1/\omega_g,1/\omega_s\right\}$, on condition that the SD decays sufficiently fast. Such short-time behavior is independent of the low-frequency structure of the SD. Instead, the long-time increasing or decreasing regime is determined by the structure of the SD near the upper cut-off frequency $\omega_g$ and by the limit $\psi\left(\infty\right)$ of the time-dependent angle $\psi(t)$ for $t\to \infty$.  Such limit coincides with the limit $\phi\left(\infty\right)$ if the conditions providing Eqs. (\ref{phiinf0})-(\ref{phiinf3}) hold. If the power $\alpha_0$ takes odd natural values and the logarithmic power $n_0$, for the first class of SDs, and $\beta_0$, for the second class, vanishes, the limit $\psi\left(\infty\right)$ depends on the power $\alpha_{k_2}$ appearing in Eq. (\ref{o0log}). The number $k_2$ is the least non-vanishing index such that either $\alpha_{k_2}$ is not an odd natural number or it is and the power $n_{k_2}$, for the first class of SDs, or $\beta_{k_2}$, for the second class, is 
non-vanishing. In fact, we find
\begin{eqnarray}
&&\hspace{-2em}\psi\left(\infty\right)=
\frac{\pi}{2} \left(2+ \left((-1)^m-1\right)\operatorname{sign}\left(\cos\left(\frac{\pi}{2}\alpha_{k_2}\right)\right)\right),
\label{psiinf4}\\
&&\hspace{-2em}\textrm{if} \hspace{1em}
\alpha_0=1+2 m, \hspace{1em}\textrm{and} \hspace{1em} \,n_0=\beta_0= 0, \nonumber
\end{eqnarray}
while, if $\alpha_{k_2}=1+2 m_2$ where $m_2$ is a natural number, the limit is 
\begin{eqnarray}
&&\hspace{-2em}\psi\left(\infty\right)=\frac{\pi}{2} \left(2
+(-1)^{m_2}\left((-1)^{m}-1\right)\right),\label{psiinf5}
\\&&\hspace{-2em}\textrm{if} \hspace{1em}\alpha_0=1+2 m,
\hspace{1em}\textrm{and} \hspace{1em}n_0=\beta_0= 0. \nonumber
\end{eqnarray}

At this stage, consider long times, $t \gg 1/\omega_s$, such that 
\begin{eqnarray}
\left|\psi(t)-\psi\left(\infty\right)\right|<\varepsilon_0, \label{Cepsilon1}
\end{eqnarray}
where $0<\varepsilon_0<\pi/2$. Such long times exist since the the function is continuous and the limit $\psi\left(\infty\right)$ exists finite. At these times, Eq. (\ref{dotEgap}) suggests that the bath energy certainly increases over the intervals $\left[t^{(+)}_{1,n},t^{(+)}_{2,n}\right]$ and certainly decreases over the intervals $\left[t^{(-)}_{1,n},t^{(-)}_{2,n}\right]$, where
\begin{eqnarray}
&&\hspace{-1.7em}t^{(+)}_{1,n}=\frac{2n\pi - \psi\left(\infty\right)+\varepsilon_0}{\omega_g},
\hspace{1em}t^{(+)}_{2,n}=\frac{\pi \left(1+2n\right)- \psi\left(\infty\right)-\varepsilon_0}{\omega_g}, \label{tpn} \\
&&\hspace{-1.7em}t^{(-)}_{1,n}=\frac{\pi \left(1+2n\right)- \psi\left(\infty\right)+\varepsilon_0}{\omega_g},
\hspace{1em}t^{(-)}_{2,n}=\frac{2\pi \left(1+n\right)- \psi\left(\infty\right)-\varepsilon_0}{\omega_g}. \label{tmn} 
\end{eqnarray}
The index $n$ takes natural values such that the instants $t^{(+)}_{1,n}$ fulfill the constraint (\ref{Cepsilon1}). For example, by choosing $n\gg\bar{n}$, where $\bar{n}=2+ \lfloor\omega_g/\omega_s\rfloor$, the condition
$t^{(+)}_{1,n}\gg 1/\omega_s$ holds for every $n\gg \bar{n}$. Consequently, every instant $t$ of the intervals $\left[t^{(+)}_{1,n},t^{(+)}_{2,n}\right]$ and $\left[t^{(-)}_{1,n},t^{(-)}_{2,n}\right]$ belongs to the long time scale, 
$t\gg 1/\omega_s$. Every time interval has the same length,
\begin{eqnarray}
&&\hspace{-1.7em}t^{(+)}_{2,n}-t^{(+)}_{1,n}=t^{(-)}_{2,n}-t^{(-)}_{1,n}=\frac{\pi-2 \varepsilon_0}{\omega_g}. \label{ltn}
\end{eqnarray}
By choosing arbitrarily small values of the positive parameter $\varepsilon_0$ the duration of each interval tends to the supremum value $\pi/\omega_g$. In such limiting conditions, the amplitudes of the oscillations of the bath energy tend to vanish. Furthermore, the intervals over which the bath energy increases tend to the following form,
\begin{eqnarray}
&&\hspace{-1.7em}\Bigg.\Bigg]\frac{2n\pi - \psi\left(\infty\right)}{\omega_g},\frac{\pi \left(1+2n\right)- \psi\left(\infty\right)}{\omega_g}\Bigg[\Bigg., \label{dtp} 
\end{eqnarray}
while the intervals over which the bath energy decreases tend to the expression below,
\begin{eqnarray}
&&\hspace{-1.7em}
\Bigg.\Bigg]
\frac{\pi \left(1+2n\right)- \psi\left(\infty\right)}{\omega_g}, 
\frac{2\pi \left(1+n\right)- \psi\left(\infty\right)}{\omega_g}\Bigg[\Bigg.. \label{dtm} 
\end{eqnarray}
Each interval is determined by the cut-off frequency $\omega_g$ and by the structure of the SD near the frequency $\omega_g$ via the limit $\psi\left(\infty\right)$.

\section{Flow of quantum information}\label{6}

The trace-distance measure introduced in Ref. \cite{BnnMarkovPRL2009} provides for the system under study a simple measure of non-Markovianity \cite{nnMarkovNeg2LZPRA2011,FPRA2013}
\begin{equation}
\mathcal{N}=\int_{\gamma(t)<0}\left|\gamma(t)\right|e^{-\Xi(t)} d t. \label{NT}
\end{equation}
A non-vanishing contribution is gained in the above integration if the dephasing rate is negative over a time interval of non-vanishing measure, and the evolution is non-Markovian. Such persistent negative values of the dephasing rate can be interpreted as information backflow from the environment in the open system. The dephasing rate is determined by the SD of the system via Eq. (\ref{gamma0}), at zero temperature, and via Eq. (\ref{gammaT}), at non-vanishing temperatures. For ohmic-like SDs, information backflow, recoherence and non-Markovian dynamics depend on the value of the ohmicity parameter \cite{MPRAr2013,MPRA2014,QbtMPRA2014,GPRA2017}. At zero temperature, $T=0$, the appearance of information backflow is facilitated by the presence of a low-frequency gap in the continuous distribution of frequency modes of the bosonic environment \cite{GgapsXiv2017}.

 Here, we investigate how the appearance of information backflow is influenced by a low-frequency gap if the external environment is initially set in a thermal state, at non-vanishing temperature $T$, which is factorized from the initial state of the qubit. The dephasing rate reads
\begin{equation}
\gamma_T(t)=\int_{\omega_g}^{\infty} \frac{J_T\left(\omega\right)}{\omega}\,\sin \left(\omega t\right) d\omega.  \label{gammaTgap}
\end{equation}
The presence of the spectral gap induces oscillations of the dephasing rate which can be studies via the equivalent form
\begin{equation}
\gamma_T(t)=\upsilon_c(t) \sin \left(\omega_g t\right)+\upsilon_s(t) \cos \left(\omega_g t\right). \label{gammaTgaposc}
\end{equation}
The functions $\upsilon_c(t)$ and $\upsilon_s(t)$ are defined 
in terms of the SD as follows,
\begin{eqnarray}
&&\hspace{-1em}\upsilon_c(t) =\int_0^{\infty}\frac{J_T\left(\omega_g+ \omega^{\prime}\right)}{\omega_g+ \omega^{\prime}}\, \cos\left(\omega^{\prime} t\right) d \omega^{\prime}, \label{uc} \\
&&\hspace{-1em}\upsilon_s(t) =\int_0^{\infty}\frac{J_T\left(\omega_g+ \omega^{\prime}\right)}{\omega_g+ \omega^{\prime}}\, \sin\left(\omega^{\prime} t\right) d \omega^{\prime}. \label{us}
\end{eqnarray}
Again, if the constraint $\upsilon^2_c(t)+\upsilon^2_s(t)>0$ is fulfilled for every $t \geq 0$, the dephasing rate reads
\begin{eqnarray}
&&\gamma_T(t)=R_3(t)
 \sin \left(\omega_g t +\xi(t)\right). \label{gammaS}
\end{eqnarray}
where $R_3(t)=\sqrt{\upsilon^2_c(t)+\upsilon^2_s(t)}$. Similarly to the angles $\phi(t)$ and $\psi(t)$, the time-dependent angle $\xi(t)$ vanishes for $t=0$, i.e., $\xi(0)=0$, and is defined for $t>0$ via the functions $\upsilon_c(t)$ and $\upsilon_s(t)$ as below,
\begin{eqnarray}
&&\hspace{-1em}\xi(t)=0,\hspace{1em}\text{if}\hspace{1em} \upsilon_s(t)=0 \hspace{1em}\text{and}\hspace{1em}\upsilon_c(t)>0,\label{xi0} \\
&&\hspace{-1em}\xi(t)=\pi,\hspace{1em}\text{if}\hspace{1em} \upsilon_s(t)=0 \hspace{1em}\text{and}\hspace{1em}\upsilon_c(t)<0,\label{xi1} \\
&&\hspace{-1em}\xi(t)=\operatorname{arccot} \frac{\upsilon_c(t)}{\upsilon_s(t)},\hspace{1em}\text{if}\hspace{1em} \upsilon_s(t)>0,\label{xi2} \\
&&\hspace{-1em}\xi(t) =\pi +\operatorname{arccot} \frac{\upsilon_c(t)}{\upsilon_s(t)},\hspace{1em}\text{if}\hspace{1em} \upsilon_s(t)<0. \label{xi3}
\end{eqnarray}
Again, in general, the time-dependent angle $\xi(t)$ is not a continuous function of time. On the contrary, the function $\xi(t)$ is continuous for every $t\geq 0$, bounded for $t \to \infty$ and differentiable for $t>0$ if the sine transform $\upsilon_s(t)$ is positive (negative), $\upsilon_s(t)>_{\left(<\right)}0$ for every $t > 0$, and if the cosine transform $\upsilon_c(t)$ and the sine transforms $\upsilon_s(t)$ are continuous for every $t \geq0$, and differentiable for every $t>0$. For example, consider regular SDs such that the corresponding functions $\upsilon_c(t)$ and $\upsilon_s(t)$ are continuous and differentiable, and such that the constraint
\begin{eqnarray}
J^{\prime}\left(\omega\right)<\left(\frac{1}{\omega}
+\frac{1}{T} \, \operatorname{cosech}\frac{\omega}{T}\right)J\left(\omega\right). \label{conSDT}
\end{eqnarray}
holds for every $\omega>\omega_g$. If such conditions are fulfilled by the SD, the sine transform $\upsilon_s(t)$ is positive, $\upsilon_s(t)>0$ for every $t > 0$ and the angle $\xi(t)$, given by Eq. (\ref{xi2}), belongs to the interval $\left.\right]0,\pi\left[\right.$. Notice that the constraint (\ref{conSDphisp}) is obtained from the inequality (\ref{conSDT}) in the limit of vanishing temperature, $T \to 0^+$.

Similarly to the case of a structured reservoir of frequency modes at zero temperature \cite{GgapsXiv2017}, we find an infinite sequence of time intervals over which information backflow appears, if $\upsilon_c^2(t)+\upsilon_s^2(t)>0$ for every $t \geq 0$, and if the angle $\xi(t)$ is a continuous function of time and the limit $\xi\left(\infty\right)$ exists finite for $t \to \infty$. Under the conditions described above, and similarly to the case of zero temperature \cite{GgapsXiv2017}, temporal bounds can be determined for the episodes of information backflow. The first episode of information backflow has already started at the instant $\pi/\omega_g$, and has already ended at the instant $2\pi/\omega_g$. The sequence of information backflows is infinite and, for every $n=1,2,\ldots$, the $n$th backflow has already started at the instant $\pi \left(1+2(n-1)\right)/\omega_g$ and has already ended at the instant $2\pi n/\omega_g$. Every time interval is no longer than $\pi/\omega_g$.

The short-time behavior of the dephasing rate is determined by the high-frequency structure and integral properties of the SD. A common short-time behavior is obtained if the SDs decay sufficiently fast at high frequencies. In fact, 
for $t \ll \min\left\{1/ \omega_s,1/\omega_g\right\}$, the dephasing rate increases linearly,
\begin{equation}
\gamma_T(t) \sim l_T t,\label{gammaTshort}
\end{equation}
if the SDs belong to the first class and $\chi_0>1$, or to the second class and $\chi_0>3$. The parameter $l_T$ is defined in the Appendix in terms of integral properties of the SD. Consequently, the information is lost over short times.

Similarly to the behavior of the dephasing rate at zero temperature \cite{GgapsXiv2017}, if the external environment is initially in a thermal state at temperature $T$ and factorized from the initial state of the qubit, the episodes of information backflow become regular over long times. Again, if the cosine transform $\upsilon_c(t)$ and the sine transform $\upsilon_s(t)$ are continuous and differentiable and if the sine transform $\upsilon_s(t)$ has constant sign over long times, $t \gg 1/\omega_s$, the long-time intervals over which information backflow appears can be described in terms of the limit $\xi\left(\infty\right)$. Under the conditions described by Eqs. (\ref{phiinf0})-(\ref{phiinf3}), the limit $\xi\left(\infty\right)$ coincides with the limits $\psi\left(\infty\right)$ and $\phi\left(\infty\right)$. On the contrary, if the power $\alpha_0$ takes odd natural values and the logarithmic power $n_0$, for the first class of SDs, and $\beta_0$, for the second class, vanishes, the limit $\xi\left(\infty\right)$ depends on the power $\alpha^{\prime \prime}_{k_3}$. Such power appears in Eq. (\ref{LsT1}) of the Appendix, for the first class of SDs, or in Eq. (\ref{LsT2}), for the second class. The natural number $k_3$ is the least non-vanishing index such that either $\alpha^{\prime \prime}_{k_3}$ is not odd, or it is odd and the logarithmic power $n^{\prime \prime}_{k_3}$, for the first class of SDs, or $\beta^{\prime \prime}_{k_3}$, for the second class, is non-vanishing. In the latter conditions we obtain \cite{GgapsXiv2017,GXiv2016}
\begin{eqnarray}
&&\hspace{-2em}\xi\left(\infty\right)=
\frac{\pi}{2} \left(2+ \left((-1)^m-1\right)\operatorname{sign}\left(\cos\left(\frac{\pi}{2}\alpha^{\prime \prime}_{k_3}\right)\right)\right),
\label{xiinf4}\\
&&\hspace{-2em}\textrm{if} \hspace{1em}
\alpha_0=1+2 m, \hspace{1em}\textrm{and} \hspace{1em} \,n_0=\beta_0= 0, \nonumber
\end{eqnarray}
where $m$ is a natural number, while, if $\alpha^{\prime \prime}_{k_3}=1+2 m_3$, where $m_3$ is a natural number, the limit is 
\begin{eqnarray}
&&\hspace{-2em}\xi\left(\infty\right)=\frac{\pi}{2} \left(2
+(-1)^{m_3}\left((-1)^{m}-1\right)\right),\label{xiinf5}
\\&&\hspace{-2em}\textrm{if} \hspace{1em}\alpha_0=1+2 m,
\hspace{1em}\textrm{and} \hspace{1em}n_0=\beta_0= 0. \nonumber
\end{eqnarray}

Consider long times, $t \gg 1/\omega_s$, such that 
\begin{eqnarray}
\left|\xi(t)-\xi\left(\infty\right)\right|<\varepsilon_0, \label{Cepsilon2}
\end{eqnarray}
where $0<\varepsilon_0<\pi/2$. Such long times exist since the 
limit $\xi\left(\infty\right)$ exists finite. At these times, the expression (\ref{gammaS}) suggests that the dephasing rate is certainly positive over the intervals $\left[t^{(+)}_{3,n},t^{(+)}_{4,n}\right]$, given by 
\begin{eqnarray}
&&\hspace{-1.7em}t^{(+)}_{3,n}=\frac{2n\pi - \xi\left(\infty\right)+\varepsilon_0}{\omega_g}, 
\hspace{1em}t^{(+)}_{4,n}=\frac{\pi \left(1+2n\right)- \xi\left(\infty\right)-\varepsilon_0}{\omega_g}. \label{tpn3}
 \end{eqnarray}
Negative values of the dephasing rate and information backflow appears over the intervals $\left[t^{(-)}_{3,n},t^{(-)}_{4,n}\right]$, where
\begin{eqnarray}
&&\hspace{-1.7em}t^{(-)}_{3,n}=\frac{\pi \left(1+2n\right)- \xi\left(\infty\right)+\varepsilon_0}{\omega_g}, 
\hspace{1em}t^{(-)}_{4,n}=\frac{2\pi \left(1+n\right)- \xi\left(\infty\right)-\varepsilon_0}{\omega_g}. \label{tmn3} 
\end{eqnarray}
Again, the index $n$ takes natural values such that the instants $t^{(+)}_{3,n}$ fulfill the constraint (\ref{Cepsilon2}). Again, by choosing $n\gg\bar{n}$, we find $t^{(+)}_{3,n}\gg 1/\omega_s$ for every $n\gg \bar{n}$. Consequently every instant $t$ of the intervals $\left[t^{(+)}_{1,n},t^{(+)}_{2,n}\right]$ and $\left[t^{(-)}_{1,n},t^{(-)}_{2,n}\right]$ belongs to the long time scale, 
$t\gg 1/\omega_s$. The length of the time interval is constant,
\begin{eqnarray}
&&\hspace{-1.7em}t^{(+)}_{4,n}-t^{(+)}_{3,n}=
t^{(-)}_{4,n}-t^{(-)}_{3,n}=\frac{\pi-2 \varepsilon_0}{\omega_g}. \label{ltn3}
\end{eqnarray}
Information is certainly lost in the environment over the intervals $\left[t^{(+)}_{3,n},t^{(+)}_{4,n}\right]$ and certinly flows back in the open systems over the intervals $\left[t^{(-)}_{3,n},t^{(-)}_{4,n}\right]$. Again, the duration of each interval tends to the supremum value $\pi/\omega_g$ for arbitrarily small values of the positive parameter $\varepsilon_0$. For such arbitrarily small values, the intervals over which information is lost tend to the form
\begin{eqnarray}
&&\hspace{-1.7em}\Bigg[\frac{2n\pi - \xi\left(\infty\right)}{\omega_g},\frac{\pi \left(1+2n\right)- \xi\left(\infty\right)}{\omega_g}\Bigg], \label{dtp3} 
\end{eqnarray}
while the intervals over which the dephasing rate is negative, and information is gained by the open system, tend to the expression below,
\begin{eqnarray}
&&\hspace{-1.7em}
\Bigg.\Bigg]
\frac{\pi \left(1+2n\right)- \xi\left(\infty\right)}{\omega_g}
,\frac{2\pi \left(1+n\right)- \xi\left(\infty\right)}{\omega_g}\Bigg[\Bigg.. \label{dtm3} 
\end{eqnarray} In such limiting conditions, the magnitude of the dephasing rate tends to vanish. Again, each interval is determined by the 
cut-off frequency $\omega_g$ and by the structure of the SD near the frequency $\omega_g$ via the limit $\xi\left(\infty\right)$.

\section{Correspondence between variations of the bath energy and flow of quantum information}\label{7}

At this stage we search for connections between the variations of the bath energy and flow of quantum information, over short and long times. The variations of the bath energy, corresponding to the special correlated initial conditions, have been analyzed in Sect. \ref{5} over short and long times, while the information flow, obtained for the factorized initial configurations, has been studied via the dephasing factor in Sect. \ref{6} over short and long times. By comparing variations of bath energy and information backflow, notice as first qualitative observation, that the presence of the spectral gap allows the identification of intervals over which energy certainly increases or decreases, for the special correlated condition, and information is certainly gained or lost by the open system, for the factorized initial configurations. In fact, if the involved sine and cosine transforms are continuous and differentiable and the sine transforms have constant sign, the bath energy increases (decreases) and information is lost (gained) over sequences of countable time intervals. In general, for every $n=1,2,\ldots$, 
same temporal bounds can be found for the $n$th interval of decreasing energy and the $n$th interval of information backflow.

As far as the short-time behaviors are concerned, $t \ll \min\left\{1/ \omega_s,1/\omega_g\right\}$, the bath energy increases and information is lost in the environment, if the SDs decay sufficiently fast over high frequencies, $\chi_0>1$ for the first class, or $\chi_0>3$ for the second class. Consequently, the spectral properties that induce an increase of the bath energy, for the special correlated initial states, provide information loss in the open system for the factorized initial configurations.

As far as the long-time behaviors are concerned, consider SDs which belong to the two classes under study and such that the power $\alpha_0$ is not odd, or it is odd and the logarithmic power $n_0$, for the first class, or $\beta_0$, for the second class, does not vanish. For such environmental spectra, consider long times, $t \gg 1/\omega_s$, such that both the constraints (\ref{Cepsilon1}) and (\ref{Cepsilon2}) are fulfilled. Over such long times, sequences of intervals are determined via the low-frequency structure of the SD, over which the bath energy certainly increases (decreases). Over such long-time intervals the information is certainly lost (gained) by the open system. Notice that such correspondence is derived from the crucial relation $\psi\left(\infty\right)=\xi\left(\infty\right)$. Essentially, such equality holds if the SDs under study are not tailored near the band gap edge according to power-law profiles with odd natural powers. For such SDs, the correspondence between increasing (decreasing) energy and information loss (gain) fails, since the limits $\psi\left(\infty\right)$ and $\xi\left(\infty\right)$ are, in general, different.

\section{Summary and conclusions}\label{8}

In local dephasing channels the variations of the bath energy, 
obtained for special correlated initial conditions, are related to the loss or gain of information by the open system, found for factorized initial configurations. Such correspondence depends on the spectral properties of the system through the SD. The correspondence is analyzed in Ref. \cite{GOSID2017} in case the SDs are ohmic-like at low frequencies and, possibly, are perturbed with logarithmic laws. This means that no gap exists in the continuous distribution of the frequency modes of the bosonic bath over low frequencies. Over short times the bath energy increases for the special correlated initial conditions, while the
information is lost in the environment for the factorized initial configurations. Over long times the bath energy 
either increases or decreases monotonically to the asymptotic value, depending on the low-frequency structure of the ohmic-like SD via the ohmicity parameter \cite{GOSID2017}. In the super-ohmic regime perfect correspondence is found over long times between the increase (decrease) of the bath energy, occurring for the special correlated initial conditions, and the gain (loss) of information by the open system, obtained for the factorized initial configurations. Such correspondence holds even if the temperatures involved in the special correlated and factorized initial conditions are different \cite{GOSID2017}. 

Information backflow appears in local dephasing channels for ohmic-like SDs uniquely if the ohmicity parameter exceeds a temperature-dependent critical value \cite{MPRAr2013,MPRA2014}. The presence of a low-frequency gap in the continuous distribution of the frequency modes of the environment facilitates the appearance of information backflow \cite{GgapsXiv2017}. In fact, at zero temperature, information backflow manifests over an infinite sequence of time intervals due to the low-frequency gap. Even if such intervals are generally irregular, upper bounds on the starting and ending time of each interval are found in terms of the upper cut-off frequency of the spectral gap. The intervals become regular over long times \cite{GgapsXiv2017}.

As a continuation of the scenario described above, here, we have considered a theoretical model of a local dephasing channel with a low-frequency gap in the continuous distribution of frequency modes. The bath energy has been evaluated for the special correlated initial conditions mentioned above, which are obtained from the thermal equilibrium of the whole system by performing a selective measure on the qubit \cite{M1,M2,M3}. The flow of quantum information has been studied if the state of the qubit is initially factorized from the thermal state of the bosonic bath, by analyzing the sign of the dephasing rate at non-vanishing temperatures \cite{nnMarkovNeg2LZPRA2011,FPRA2013,MPRAr2013,MPRA2014,QbtMPRA2014}. 
The absence of bosonic modes below the upper cut-off frequency of the spectral gap, and the initial conditions, either specially correlated or factorized, are crucial features for the present analysis. Both the bath energy and the dephasing rate show oscillations which are generally irregular. Still, the intervals over which the bath energy increases (decreases) exhibit the same upper bounds as those over which the dephasing rate is positive (negative), if the SDs fulfill appropriate constraints.

According to the present analysis, the bath energy increases and the information is lost by the open system in the environment, over short times, if the SD decays sufficiently fast at high frequencies. Same short-time correspondence holds without the low-frequency gap \cite{GOSID2017}. Over long times, the bath energy exhibits regular damped oscillations around the asymptotic value, showing alternate increasing and decreasing behaviors. The frequency of the asymptotic oscillations coincides with the upper cut-off frequency of the spectral gap. For the factorized initial conditions and at non-vanishing temperatures the dephasing rate exhibits an oscillatory behavior that is similar to the one shown by the bath energy for the special correlated initial configurations. Except for particular SDs, at sufficiently long times, sequences of regular time intervals are found over which the bath energy increases (decreases) and the open quantum system looses (gains) information. 
The correspondence between variations of the bath energy and of information in the open system holds even for different values of the temperatures which are involved in the initial conditions. Qualitatively, under the mentioned initial conditions, the spectral gap induces, and regularizes over long times, the oscillatory behaviors of the bath energy and of the dephasing rate. Furthermore, over the determined long-time intervals, the bath energy varies inversely with respect to information which is gained or lost by the open system. The correspondence fails if the SDs are tailored near the upper cut-off frequency of the spectral gap as power laws with odd natural powers. Notice that, without the low-frequency gap, the long-time correspondence between variations of the bath energy, which either increases or decreases monotonically, and information, which is uniquely gained or lost by the open system, is reversed. In conclusion, the present theoretical construct shows how the bath energy and the information flow in local dephasing channels can be regularized, controlled and correlated, over long times, by introducing a spectral gap in the low-frequency distribution of bosonic modes and by preparing special correlated and factorized initial conditions which involve thermal states.

\appendix\label{A}
\section{Details}

For the special correlated initial conditions given by Eqs. (\ref{rho0corr}) and (\ref{rhoE0}), the energy of the bosonic bath is described by Eq. (\ref{Etosc}). Due to the low-frequency gap, the function $\Pi(t)$ is given by Eq. (\ref{Pitgapcs}) and is equivalent to Eq. (\ref{Pitgap}), if $\varphi^2_c(t)+\varphi^2_s(t)>0$ for every $t \geq 0$. The time-dependent angle $\phi(t)$ is obtained via straightforward trigonometric relations \cite{GradRyz} between the functions $\operatorname{arccot}(x)$ and $\sin(x)$ or $\cos(x)$. The cosine transform $\varphi_c(t)$ and sine transform $\varphi_s(t)$ are studied in terms of the dimensionless functions $f_{c,0}\left(\tau\right)$ and $f_{s,0}\left(\tau\right)$, which are defined as
\begin{eqnarray}
&&\hspace{-1em}f_{c,0}\left(\tau\right) =
\int_0^{\infty}\Lambda_0\left(\nu\right)\, \cos\left(\nu \tau\right) d \nu, 
\label{phctau} \\ &&   \hspace{-1em} f_{s,0}(t) =
\int_0^{\infty}\Lambda_0\left(\nu\right)\, \sin\left(\nu \tau\right) d \nu, 
\label{phstau}
\end{eqnarray}
where $\tau=\omega_s t$. In this way, the functions $\varphi_c(t)$ and $\varphi_s(t)$ read
$\varphi_c(t)=\omega_s f_{c,0}\left(\tau\right)/\nu_0$ and 
$\varphi_s(t)=\omega_s f_{s,0}\left(\tau\right)/\nu_0$, where $\nu_0=\omega_g/\omega_s$. The function $\Lambda_0\left(\nu\right)$ is defined according to Eqs. (\ref{phc}) and (\ref{phs}) for every 
$\nu\geq 0$ in terms of the
auxiliary function $\Omega\left(\nu\right)$ as
\begin{eqnarray}
\Lambda_0\left(\nu\right)=\frac{\Omega\left(\nu\right)}{1+ \nu/\nu_0}. 
\label{Lambda0}
\end{eqnarray}
The asymptotic behavior of the function $\Lambda_0\left(\nu\right)$ as $\nu\to 0^+$ 
is given by 
\begin{eqnarray}
\hspace{-2em}
\Lambda_0\left(\nu\right)\sim\sum_{r=0}^{\infty}
\sum_{k=0}^{n^{\prime}_r}c^{\prime}_{r,k} \nu^{\alpha^{\prime}_r}
\left(- \ln \nu\right)^k, \label{Ls01}
\end{eqnarray}
for the first class of SDs, and by
\begin{eqnarray}
\hspace{-2em}
\Lambda_0\left(\nu\right)\sim\sum_{r=0}^{\infty}
w^{\prime}_{r} \nu^{\alpha^{\prime}_r}\left(- \ln \nu\right)^{\beta^{\prime}_r}, \label{Ls02}
\end{eqnarray}
for the second class. Due to the properties of the powers $\alpha_j$ reported in Sect. \ref{4}, the following relations hold, $\alpha_{0}=\alpha_{0}^{\prime}$, 
$\alpha_{r+1}^{\prime}>\alpha_{r}^{\prime}$ for every $r=0,1,2,\ldots$, 
$\alpha_{r}^{\prime}\uparrow \infty$ as $r\to\infty$, for both the classes of SDs. Also, we find $n_0=n^{\prime}_0$ and 
$c_{0,k}=c^{\prime}_{0,k}$ for every $k=0,\ldots,n_0$, for the first class of SDs,
 and $\beta_0=\beta^{\prime}_0$ and $w_0=w^{\prime}_0$, for the second class.

The behaviors of the functions $f_{c,0}\left(\tau\right)$ and $f_{s,0}\left(\tau\right)$ for $\tau \ll 1$ and $\tau \gg 1$ are analyzed in Refs. \cite{GgapsXiv2017,GXiv2016} in terms of the asymptotic behaviors of the function $\Lambda_0\left(\nu\right)$. 
The function $\Lambda_0\left(\nu\right)$ must fulfill the constraints which are required for the auxiliary function 
$\Omega\left(\nu \right)$ in Sect. \ref{4}. See Refs. \cite{GgapsXiv2017,GXiv2016} for details. The asymptotic analysis performed in such references holds also for $0\geq\alpha_0>-1$. Following the results of Ref. \cite{GgapsXiv2017,GXiv2016}, we obtain that if the SDs belong to the first class and $\chi_0>1$, or to the second class and $\chi_0>3$, the functions $\varphi_c(t)$ and $\varphi_s(t)$ evolve algebraically for $t \ll 1/\omega_s$,
\begin{eqnarray}
&&\varphi_c(t)\sim l_{0,0}-l_{0,2} t^2,\hspace{1em}\varphi_s(t)\sim l_{0,1} t, \label{phscshort}
\end{eqnarray}
where
\begin{eqnarray}
&&l_{0,0}= \int_{\omega_g}^{\infty}\frac{J\left(\omega\right)}{\omega}\, d\omega, \hspace{1em} l_{0,1}= \int_{\omega_g}^{\infty}\frac{J\left(\omega\right)}{\omega}\,\left(\omega-\omega_g\right) d\omega, \nonumber \\
&&l_{0,2}= \frac{1}{2}\int_{\omega_g}^{\infty}\frac{J\left(\omega\right)}{\omega}\,\left(\omega-\omega_g\right)^2 d\omega.  \nonumber 
\end{eqnarray}
In this way, Eq. (\ref{Etshort}) is obtained, where 
\begin{eqnarray}
&&l_E=d_0\left(\frac{1}{2}\,\omega_g^2 l_{0,0}+\omega_g  l_{0,1}
+l_{0,2} \right). \nonumber
\end{eqnarray}
Following Refs. \cite{GXiv2016,GgapsXiv2017}, the long-time behaviors of the functions $\varphi_c(t)$ and $\varphi_s(t)$ are found in terms of the powers $\alpha^{\prime}_r$ and $n^{\prime}_r$, for the first class of SDs, or $\beta^{\prime}_r$, for the second class, which are defined via Eqs. (\ref{Ls01}) and (\ref{Ls02}). The property $\alpha_{0}=\alpha_{0}^{\prime}$ is crucial for the analysis of the variations of the bath energy which is performed in Sect. \ref{5}. The limit $\phi \left(\infty\right)$ is given by Eqs. (\ref{phiinf0})-(\ref{phiinf5}), via the asymptotic behaviors of the functions $\varphi_c(t)$ and $\varphi_s(t)$. Refer to \cite{GXiv2016,GgapsXiv2017} for details. The expression (\ref{Etosc}) of the energy, Eq. (\ref{Pitgap}), the continuity of the function $\phi(t)$ and the existence of the limit $\phi\left(\infty\right)$ lead to Eqs. (\ref{EtAsympt1}) and (\ref{EtAsympt2}) via the constraint (\ref{Cepsilon0}).

 By applying the procedure described above to the analysis of the derivative of the bath energy, $\dot{\epsilon}_E(t)$, we obtain Eqs. (\ref{dotEgap})-(\ref{psi3}). The regimes of increasing (decreasing) energy correspond to positive (negative) values of the time derivative $\dot{\epsilon}_E(t)$ and are obtained from Eq. (\ref{dotEgapS}). The energy increases over times $t$ such that
\begin{equation}
2 n \pi<\omega_g t+\psi(t)<\pi(1+2 n),
\label{posang1}
\end{equation}
and decreases over times $t$ such that 
\begin{equation}
\pi \left(1+2n\right)<\omega_g t+\psi(t)<2\pi(1+n),
\label{negang1}
\end{equation}
for every $n=0,1,2,\ldots$. The above relations and the constraint (\ref{Cepsilon1}) define the long-time intervals over which the bath energy either increases or decreases, respectively. Under the conditions providing Eqs. (\ref{phiinf0})-(\ref{phiinf3}), the limits $\phi\left(\infty\right)$ and $\psi\left(\infty\right)$ coincide. If the power $\alpha_0$ takes odd natural values and the logarithmic power $n_0$, for the first class of SDs, or $\beta_0$, for the second class, vanishes, the limit $\psi\left(\infty\right)$ is given by Eqs. (\ref{psiinf4}) and (\ref{psiinf5}). See Ref. \cite{GXiv2016} for details. Once the limit $\psi\left(\infty\right)$ is determined, Eq. (\ref{tpn}) describes the long-time intervals over which the bath energy certainly increases, while Eq. (\ref{tmn}) provides the long-time intervals over which the bath energy certainly decreases. Notice that the bath energy can increase outside the intervals given by Eq. (\ref{tpn}), and decrease outside the intervals given by Eq. (\ref{tmn}). Still, over such intervals, the increasing or decreasing regimes are determined. The limiting condition $\varepsilon_0\to 0^+$ provides the asymptotic intervals given by Eq. (\ref{dtp}), over which the bath energy increases, and the asymptotic intervals described by Eq. (\ref{dtm}), over which the bath energy decreases.

At non-vanishing temperatures the dephasing rate, defined by Eq. (\ref{gammaT}), is given by Eq. (\ref{gammaTgap}), due to the presence of the spectral gap. The latter expression is analyzed via the change of variable $\omega=\omega_g + \omega^{\prime}$. In this way, the Eqs. (\ref{gammaTgaposc}) and (\ref{gammaS}) are obtained. The functions $\upsilon_c(t)$ and $\upsilon_s(t)$, defined by Eqs. (\ref{uc}) and (\ref{us}), are analyzed by introducing the dimensionless functions $f_{c,T}\left(\tau\right)$ and $f_{s,T}\left(\tau\right)$. Such functions are defined via the relations $\upsilon_c(t)=\omega_s f_{c,T}\left(\tau\right)$ and 
$\upsilon_s(t)=\omega_s f_{s,T}\left(\tau\right)$, and are given by
\begin{eqnarray}
&&\hspace{-1em}f_{c,T}\left(\tau\right) = \int_0^{\infty}\Lambda_T\left(\nu\right) \cos\left(\nu \tau\right) d\nu, \label{varphic1ad} \\
&&\hspace{-1em}f_{s,T}\left(\tau\right) = \int_0^{\infty}\Lambda_T\left(\nu\right)\, \sin\left(\nu \tau\right) d\nu,
\label{varphis1ad}
\end{eqnarray}
where $\tau=\omega_s t$. The dimensionless function $\Lambda_T\left(\nu\right)$ is defined in terms of the auxiliary function $\Omega\left(\nu\right)$ as 
\begin{eqnarray}
\Lambda_T\left(\nu\right)=Q_T\left(\nu\right) \Omega\left(\nu\right). \label{LambdaT}
\end{eqnarray}
 The function $Q_T\left(\nu\right)$ is defined by the form
\begin{eqnarray}
Q_T\left(\nu\right)=\frac{q_0 }{1+\nu/\nu_0}\, \frac{1+\left(\tanh \left(\nu/\nu_1\right)\right)/q_1}{1+ q_1 \tanh \left(\nu/\nu_1\right)}, \label{QT}
\end{eqnarray}
where $\nu_1=2T/ \omega_s$, $q_0=q_1/\nu_0$ and $q_1=\operatorname{coth}\left(\nu_0/\nu_1\right)$. The function $Q_T\left(\nu\right)$ admits a power series expansion in $\nu=0$, and the auxiliary function $\Omega\left(\nu\right)$ behaves as $\nu\to 0^+$ according to Eq. (\ref{o0log}), for the first class of SDs, and according to Eq. (\ref{OmegaLog0}) for the second class. Consequently, 
the function $\Lambda_T\left(\nu\right)$ behaves for $\nu\to 0^+$ as
\begin{eqnarray}
\hspace{-2em}
\Lambda_T\left(\nu\right)\sim\sum_{r=0}^{\infty}
\sum_{k=0}^{n^{\prime\prime}_r}c^{\prime\prime}_{r,k} \nu^{\alpha^{\prime\prime}_r}\left(- \ln \nu\right)^k, \label{LsT1}
\end{eqnarray}
for the first class of SDs, and as 
\begin{eqnarray}
\hspace{-2em}
\Lambda_T\left(\nu\right)\sim\sum_{r=0}^{\infty}
w^{\prime \prime}_{r} \nu^{\alpha^{\prime\prime}_r}\left(- \ln \nu\right)^{\beta^{\prime\prime}_r}, \label{LsT2}
\end{eqnarray}
for the second class. Due to the properties of the powers $\alpha_j$, the following relations hold for the first class of SDs, $\alpha_{0}=\alpha_{0}^{\prime\prime}$, $\alpha_{r+1}^{\prime\prime}>\alpha_{r}^{\prime\prime}$ for every $r=0,1,2,\ldots$, $\alpha_{r}^{\prime\prime}\uparrow \infty$ as $r\to\infty$, $n_0=n^{\prime\prime}_0$ and $c^{\prime\prime}_{0,k}=q_0 c_{0,k}$ for every $k=0,\ldots,n_0$. For the second class of SDs, we find $\alpha_{0}=\alpha_{0}^{\prime\prime}$, $\alpha_{r+1}^{\prime\prime}>\alpha_{r}^{\prime\prime}$ for every $r=0,1,2,\ldots$, $\alpha_{r}^{\prime\prime}\uparrow \infty$ as $r\to\infty$, $\beta_0=\beta^{\prime\prime}_r$ and $w^{\prime\prime}_0=q_0 w_0$. The property $\alpha_{0}=\alpha_{0}^{\prime\prime}$ is crucial for the analysis of the information flow, which is performed in Sect. \ref{6}, and for the comparison with the variations of the bath energy, which is performed in Sect. \ref{7}.

The short-time behavior of the functions $f_{c,T}$ and $f_{s,T}$ is determined by integral and asymptotic properties of the function $\Lambda_T\left(\nu\right)$. The function $\Lambda_T\left(\nu\right)$ must fulfill the constraints which are required in Sect. \ref{4} for the auxiliary function $\Omega \left(\nu\right)$. Refer to \cite{GXiv2016} for details. The analysis performed in such reference holds also for $0\geq\alpha_0>-1$. Notice that the logarithmic power $n_0$, for the first class of SDs, or $\beta_0$, for the second class, must vanish, $n_0=\beta_0=0$, in case $\alpha_0=0$, due to the constraint of summability. 
Following the results of Ref. \cite{GXiv2016}, we find that if the SDs belong to the first class and $\chi_0>1$, or to the second class and $\chi_0>3$, the functions $\upsilon_c(t)$ and $\upsilon_s(t)$ evolve algebraically for $t \ll 1/\omega_s$,
\begin{eqnarray}
&&\upsilon_c(t)\sim l_{T,0}-l_{T,2} t^2,\hspace{1em}\upsilon_s(t)\sim l_{T,1} t, \label{phscshort}
\end{eqnarray}
where
\begin{eqnarray}
&&l_{T,0}= \int_{\omega_g}^{\infty}\frac{J_T\left(\omega\right)}{\omega}\, d\omega, \hspace{1em} l_{T,1}= \int_{\omega_g}^{\infty}\frac{J_T\left(\omega\right)}{\omega}\,\left(\omega-\omega_g\right) d\omega, \nonumber \\
&&l_{T,2}= \frac{1}{2}\int_{\omega_g}^{\infty}\frac{J_T\left(\omega\right)}{\omega}\,\left(\omega-\omega_g\right)^2 d\omega.  \nonumber 
\end{eqnarray}
In this way, Eq. (\ref{gammaTshort}) is obtained, where $l_T= \omega_g l_{T,0}+l_{T,1}$. Following Ref. \cite{GXiv2016}, the long-time behaviors of the functions $\upsilon_c(t)$ and $\upsilon_s(t)$ are found in terms of the powers $\alpha^{\prime \prime}_r$ and $n^{\prime \prime}_r$, for the first class of SDs, or $\beta^{\prime \prime}_r$, for the second class, which are defined via Eqs. (\ref{LsT1}) and (\ref{LsT2}).

Again, if $\upsilon^2_s(t)+\upsilon^2_c(t)>0$ for every $t\geq 0$, the expression (\ref{gammaS}) is obtained from Eq. (\ref{gammaTgaposc}) via straightforward trigonometric relations \cite{GradRyz}. The limit $\xi \left(\infty\right)$ of the time-dependent angle $\xi(t)$ for $t \to \infty$ is obtained from the long-time behaviors of the functions $\upsilon_c(t)$ and $\upsilon_s(t)$ in the same way as the angles $\phi(t)$ and $\psi(t)$ are evaluated. The time-dependent sign of the dephasing rate is obtained from Eq. (\ref{gammaS}). The dephasing rate is positive over times $t$ such that
\begin{equation}
2\pi n <\omega_g t+\xi(t)<\pi(1+2 n),
\label{posang2}
\end{equation}
and negative over times $t$ such that
\begin{equation}
\pi \left(1+2n\right)<\omega_g t+\xi(t)<2\pi(1+n),
\label{negang2}
\end{equation}
for every $n=0,1,2,\ldots$. The above relations and the constraint (\ref{Cepsilon2}) define the long-time intervals over which the dephasing rate is positive or negative, respectively. Under the conditions providing Eqs. (\ref{phiinf0})-(\ref{phiinf3}), the limits $\phi\left(\infty\right)$, $\psi\left(\infty\right)$ and $\xi\left(\infty\right)$ coincide. If the power $\alpha_0$ takes odd natural values and the logarithmic power $n_0$, for the first class of SDs, or $\beta_0$ for the second class, vanishes, the limit $\xi\left(\infty\right)$ is given by Eqs. (\ref{xiinf4}) and (\ref{xiinf5}). The index $k_3$, the power $\alpha_{k_3}^{\prime \prime}$ and the logarithmic power $n_{k_3}^{\prime \prime}$, for the first class of SDs, and $\beta_{k_3}^{\prime \prime}$, for the first class, appear in Eq. (\ref{LsT1}) or Eq. (\ref{LsT2}), respectively, and are defined in Sect. \ref{6}. See Ref. \cite{GXiv2016} for details. Once the limit $\xi\left(\infty\right)$ is determined, Eq. (\ref{tpn3}) describes the long-time intervals over which the dephasing rate certainly takes positive values, while Eq. (\ref{tmn3}) provides the long-time intervals over which the dephasing rate is certainly negative. Notice that positive values of the dephasing rate can appear outside the intervals given by Eq. (\ref{tpn3}), and negative values can manifest outside the intervals given by Eq. (\ref{tmn3}). Still, over such intervals, the sign of the dephasing rate and the direction of the information flow are determined. The limiting condition $\varepsilon_0\to 0^+$ provides the asymptotic intervals given by Eq. (\ref{dtp3}), over which information is lost by the open system, and the asymptotic intervals described by Eq. (\ref{dtm3}), over which information backflow appears. This concludes the demonstration of the present results.

\end{document}